\documentclass{emulateapj}
\usepackage{natbib}
\usepackage{longtable}
\usepackage{epsfig}
\usepackage{amssymb}

\newcommand{\kms}{km\,s$^{-1}$}

\newcommand{\be}{\begin{equation}}
\newcommand{\ee}{\end{equation}}

\def\mseld{M$_{*,{\rm Ein}}^{\rm LD}$}
\def\msesed{M$_{*,{\rm Ein}}^{\rm SPS}$}

\slugcomment{submitted to ApJ}

\shorttitle{The IMF of elliptical galaxies}
\shortauthors{Treu et al.\ }

\begin{document}

\title{The initial mass function of early-type galaxies\altaffilmark{1}}

\author{Tommaso Treu\altaffilmark{2,}\altaffilmark{3}, Matthew~W. Auger\altaffilmark{2}, L\'{e}on~V.~E. Koopmans\altaffilmark{4},
Rapha\"el Gavazzi\altaffilmark{2,}\altaffilmark{5}, Philip~J. Marshall\altaffilmark{2}, Adam S.~Bolton\altaffilmark{6,7}}

\altaffiltext{1}{ Based on observations made with the NASA/ESA Hubble
Space Telescope, obtained at the Space Telescope Science Institute,
which is operated by the Association of Universities for Research in
Astronomy, Inc., under NASA contract NAS 5-26555.  These observations
are associated with programs \#10174,\#10587, \#10886, \#10494,
\#10798, \#11202.}  \altaffiltext{2}{Department of Physics, University
of California, Santa Barbara, CA 93106, USA ({\tt tt@physics.ucsb.edu,
mauger@physics.ucsb.edu, pjm@physics.ucsb.edu})}
\altaffiltext{3}{Sloan Fellow; Packard Fellow}
\altaffiltext{4}{Kapteyn Institute, P.O. Box 800, 9700AV Groningen,
The Netherlands ({\tt koopmans@astro.rug.nl})}
\altaffiltext{5}{Institut d'Astrophysique de Paris, UMR7095 CNRS -
Universit\'e Paris 6, 98bis Bd Arago, 75014 Paris, France ({\tt
gavazzi@iap.fr})} \altaffiltext{6}{Beatrice Watson Parrent Fellow,
Institute for Astronomy, University of Hawai`i, 2680 Woodlawn Dr.,
Honolulu, HI 96822 ({\tt bolton@ifa.hawaii.edu})}
\altaffiltext{7}{Department of Physics and Astronomy, University of
Utah, 115 South 1400 East, Salt Lake City, UT 84112 USA
({\tt bolton@physics.utah.edu})}

\begin{abstract}
We determine an absolute calibration of the initial mass function
(IMF) of early-type galaxies, by studying a sample of 56 gravitational
lenses identified by the SLACS Survey. Under the assumption of
standard Navarro, Frenk \& White dark matter halos, a combination of
lensing, dynamical, and stellar population synthesis models is used to
disentangle the stellar and dark matter contribution for each lens. We
define an ``IMF mismatch'' parameter $\alpha\equiv$\mseld$/$\msesed as
the ratio of stellar mass inferred by a joint lensing and dynamical
models (\mseld) to the current stellar mass inferred from stellar
populations synthesis models (\msesed). We find that a Salpeter IMF
provides stellar masses in agreement with those inferred by lensing
and dynamical models ($\langle \log \alpha\rangle=0.00\pm
0.03\pm0.02$), while a Chabrier IMF underestimates them ($\langle \log
\alpha\rangle=0.25 \pm 0.03\pm0.02$).  A tentative trend is found, in
the sense that $\alpha$ appears to increase with galaxy velocity
dispersion. Taken at face value, this result would imply a non
universal IMF, perhaps dependent on metallicity, age, or abundance
ratios of the stellar populations.  Alternatively, the observed trend
may imply non-universal dark matter halos with inner density slope
increasing with velocity dispersion. While the degeneracy between the
two interpretations cannot be broken without additional information,
the data imply that massive early-type galaxies cannot have both a
universal IMF and universal dark matter halos.

\end{abstract}


\section{Introduction}

The initial mass function (IMF) is a fundamental characteristic of a
simple stellar population. Measuring the IMF from resolved stellar
populations in the local universe has been a major astrophysical
problem for decades \citep[e.g.,][]{Sal55,Cha03}.

From the point of view of understanding distant unresolved stellar
populations, the IMF holds the key to interpreting observables such as
colors and their evolution in terms of star formation history and
chemical enrichment history. Among the many astrophysical problems
where the IMF plays a key role, in this paper we will focus on the
determination of the stellar mass of galaxies from the comparison of
stellar populations synthesis models \citep[e.g.,][]{B+C03,Mar05} with
broad band colors. Stellar masses derived in this manner are often
used to construct stellar mass functions and to study the demographics
of galaxies over cosmic time \citep[e.g.,][]{B+E00,Fon++06,Bun++06}.

Unfortunately, the form of the IMF is degenerate with the derived
stellar masses, since the luminosity is typically dominated by stars
in a relatively narrow mass range. To facilitate comparison between
results of different groups, it is therefore common practice to {\it
assume} a standard, universal, and non-evolving IMF. Popular choices
are the so-called \citet{Sal55} and \citet{Cha03} IMFs.  However, this
strategy leaves behind a systematic error that is believed to be of
the order of a factor of two in stellar mass. Furthermore, the error
need not correspond to a global error on the mean, but it could depend
on the conditions of star formation and therefore for example on
galaxy type, environment, metallicity and star formation epoch.
Recent work, for example, has called into question the universality of
the IMF and suggested it may be evolving with cosmic time
\citep[e.g.,][]{vDo08,Dav08}.

The goal of this paper is to determine the absolute normalization of
stellar masses for early-type galaxies, its scatter from object to
object, and its dependence on secondary parameters such as stellar
velocity dispersion or stellar mass, by combining three independent
probes of mass. Previous works have attempted to do this by comparing
stellar masses determined from stellar populations synthesis models
with those inferred from gravitational lensing \citep[e.g.,][]{FSB08}
or stellar kinematics \citep[e.g.,][]{Pad++04,Cap++06}. However, the
combination of the two latter techniques is particularly powerful as
it allows one to reduce many of the degeneracies and disentangle the
stellar and dark component \citep[e.g.,][]{T+K04}.

To meet our goal we exploit the large and homogeneous sample of strong
gravitational lenses discovered by the Sloan Lenses ACS Survey
\citep{Bol++06, Tre++06, Koo++06, Gav++07, Bol++08a, Gav++08,
Bol++08b,Aug++09a}. These papers showed that the SLACS lenses are
statistically indistinguishable within the current level of
measurement errors from control samples in terms of size, luminosity,
surface brightness, location on the fundamental plane, environment,
stellar and halo mass. Thus our results can be generalized to the
overall population of early-type galaxies.

Recently, \citet{Gri++09} used ground based photometry to derive
stellar masses for a subset of SLACS lenses. They compared the
inferred stellar mass fractions with average lensing stellar mass
fractions determined by \citet{Koo++06} and \citet{Gav++07} for
subsamples of the SLACS lenses, finding a general agreement. The
analysis presented here takes several steps forward: i) space-based
HST photometry \citep{Aug++09a} extending into the near infrared is
used for stellar masses; ii) individual stellar fraction estimates
from lensing and dynamical models are used for each galaxy; iii) a
Bayesian framework is adopted to determine the IMF normalization and
errors for each galaxies. This progress allows us to investigate for
the first time the scatter of the IMF and its possible dependency on
galaxy velocity dispersion and hence, e.g., redshift of formation of
the stellar populations or metallicity.

We assume a concordance cosmology with matter and dark energy density
$\Omega_m=0.3$, $\Omega_{\Lambda}=0.7$, and Hubble constant
H$_0$=100$h$kms$^{-1}$Mpc$^{-1}$, with $h=0.7$ when necessary. Base-10
logarithms are used.

\section{Sample and Data}

\label{sec:data}

The sample analyzed in this paper is composed of 56 of the 58
early-type lens galaxies identified by the SLACS Survey, for which a
joint lensing and dynamical analysis has been performed
(\S~\ref{ssec:2comp}), following the methods described by
\citet{T+K04} and \citet{Koo++06}. Two of the lenses of the 
\citet{Koo++09} sample have been excluded because the 
available single-band HST photometry is not sufficient to determine
reliable stellar masses \citep{Aug++09a}. All lenses are classified as
definite (grade ``A'') based on the identification of multiple images
and are successfully modeled as a single mass component.  A full
description of the SLACS Survey and the selection process --- together
with Hubble Space Telescope images and measured photometric and
spectroscopic parameters of all the lenses --- is given in the SLACS
papers
\citep{Bol++06,Bol++08a,Aug++09a}.

\subsection{Dark and luminous mass from lensing and stellar kinematics}
\label{ssec:2comp}

The lensing models provides a very accurate and precise measurement of
the mass contained within the Einstein Radius (M$_{\rm Ein}$). As
discussed at length by, e.g., \citet{T+K04} and \citet{Koo++06}, the
addition of stellar velocity dispersion information allows one to
disentangle stellar and dark matter, given a choice of mass models
describing the dark matter halo, the stellar component and orbital
anisotropy.  In practice, the joint lensing and dynamical analysis can
be decoupled because the uncertainty on the lensing measurement is
negligible with respect to that associated to the stellar velocity
dispersion \citep[see][for joint analysis and detailed discussion of
the methodology]{Bar++09}. We thus proceed as follows. First, we
determine the total mass within the Einstein radius by fitting the
lensing geometry with a gravitational lens model. The inferred lensing
mass is determined to a few percent precision, independent of the
specific form chosen to describe the gravitational potential of the
deflector \citep[assumed to be a singular isothermal ellipsoid by
SLACS;][]{Bol++08a,Aug++09a}. We then compute the likelihood of a
family of two-component mass models (stellar and dark matter) with
respect to the stellar velocity dispersion measured by SDSS and
determine the range of acceptable solutions by computing the posterior
probability distribution function. Confidence intervals on each
individual parameter can be obtained by marginalizing over the other
parameters.

For the purpose of this paper, we adopt the so called NFW
\citep{NFW96,NFW97} model for the dark matter halo:

\begin{equation}
  \rho_{\rm DM}(r/r_b)=\frac{\rho_{\rm DM,0}}{(r/r_{\rm
  b})(1+(r/r_{\rm b}))^2}.
  \label{eq:DM}
\end{equation}

In accordance with the CDM picture (e.g. Bullock et al.\ 2001,
Macci\'o et al. 2007) we expect the break radius $r_b$ to be much
larger than the effective and Einstein radii. This makes the results
insensitive to the precise choice of $r_b$. We fix $r_b=30$ kpc in
agreement with the expected value for the average virial mass of the
SLACS sample ($\sim 10^{13}$ M$_{\odot}$; Gavazzi et al.\ 2007).

The luminous component is described as either a \citet{Jaf83} or a
\citet{Her90} model, which are good simple analytic descriptions of
the light profile of early-type galaxies, and bracket the inner slope
of the \citet{dev48} profile. The orbital anistropy of the stars is
modeled as a constant $\beta$

\begin{equation}
\beta\equiv 1-\frac{\sigma^2_{\theta}}{\sigma^2_{r}},
\label{eq:constiso}
\end{equation}

\noindent
where $\sigma_{\theta}$ and $\sigma_{r}$ are the tangential and radial
component of the pressure tensor. For clarity, we will adopt results
obtained with Hernquist-isotropic models as our default (i.e., with
$\beta=0$). As we will discuss further below, none of the results of
this paper is changed if Jaffe or moderately anisotropic models
($\beta=\pm0.25$)-- consistent with independent constraints on
anisotropy \citep[e.g.][]{Ger++01} -- are considered instead.

Since the mass within the Einstein radius is fixed by the lensing
geometry-- for a given anisotropy and functional form of the stellar
component-- the model has just one free parameter: the fraction of
stellar mass $f_*$ inside the cylinder of radius equal to the
circularized Einstein radius. Thus, for each lens, the lensing and
dynamical analysis produces the full posterior distribution function
of $f_*$, p($f_*$), assuming a uniform prior in the interval
[0,1]. The product of M$_{\rm Ein}$ and $f_*$ provides the
lensing+dynamical measurement of the stellar mass inside the Einstein
Radius \mseld for each lens galaxy. We note that p($f_*$), and thus
p(\mseld) are typically fairly asymmetric, because of the physical
requirement that the stellar mass fraction be less or equal to unity
imposed by the prior. For illustration purposes we use the median as
our best estimator of the quantity, but we use the full distribution
throughout the analysis. The average median value of $f_*$ is 80\%
with a dispersion of 17\%, consistent with the fact that most of the
mass inside the Einstein Radius, corresponding on average to half the
effective radius for the SLACS lenses, is accounted for by the stellar
component.

The main goal of this paper is to explore the constraints on the IMF
that can be gathered by {\it assuming} a standard universal NFW
profile for the dark matter halo.  More general forms of the dark
matter halo profile could also be considered with our formalism,
including for example the so-called generalized NFW profile, where the
inner slope is allowed to be a free parameter. The gNFW profiled
includes steeper profiles as well as profiles with a constant inner
core, which are believed to be appropriate for some spiral and low
surface brightness galaxies \citep[e.g.,][and references
therein]{Sal++07} as well as some clusters of galaxies \cite[e.g.][and
references therein]{San++08,New++09}. In general, allowing the inner
slope to be a free parameter results in a degeneracy between stellar
mass fraction and inner slope \citep[e.g.,][]{T+K04}, in the sense
that steeper inner slopes require less stellar mass to obtain the same
stellar velocity dispersion. The degeneracy is best reduced by
spatially resolved velocity dispersion measurements
\citep[e.g.,][]{T+K04}.  However, even with a single stellar velocity
dispersion measurement, interesting limits on the inner slope and
stellar mass fraction can be obtained by marginalizing over the other
parameter with an appropriate prior.  A full analysis of the inner
slope of dark matter halos is left for future work when more accurate
and spacially resolved velocity dispersion measurements will be
available to better constrain the inner slope (e.g., Barnab\'e et al,
2010, in preparation). However, the results presented in this paper do
not change significantly if a gNFW halo with uniform prior on the
inner slope is considered instead of a simple NFW.

\subsection{Stellar mass from stellar populations synthesis models}

The second fundamental ingredient of this work is the posterior
distribution function for the stellar mass derived by Auger et al.\
(2009) by applying stellar populations synthesis models to multicolor
HST photometry. In this paper, we consider as our reference stellar
masses based on the Bruzual \& Charlot (2003) models using an
informative prior on metallicity taken from the spectroscopic study of
Gallazzi et al.\ 2005 (hereafter the Gallazzi prior; see Auger et al.\
2009 for details). To check for possible systematics, we also consider
stellar masses based on an ``ignorant'' uniform prior on metallicity
(see Auger et al. 2009 for details), and based on stellar populations
synthesis models by \citet{Mar05}. Finally, we consider two baseline
choices of the IMF: the \citet{Cha03} and \citet{Sal55} models.

Taking into account the fraction of light inside the cylinder, this
method provides an independent measurement of the stellar mass inside
the Einstein radius \msesed. It is important to emphasize that the
current stellar mass is significantly lower than the stellar mass at
zero-age, due to mass loss during stellar evolution. For a single
stellar population with Chabrier IMF, $\sim$50\% of the initial mass
is in the form of gas at 10 billions years of age. The fraction is
significantly smaller for a Salpeter IMF ($\sim$30\%), or if different
prescriptions for stellar mass loss are adopted, as discussed, e.g.,
by \citet{Mar05}. Although the current stellar mass is the standard
quantity for this kind of analysis it is clear that our joint lensing
and dynamics analysis is sensitive to all the mass. Therefore, if a
fraction of baryons lost during stellar evolution were to retain the
phase space distribution of their parent stars, they would also be
counted by the lensing and dynamical analysis towards the component
distributed as light. Most of the residual gas in elliptical galaxies
is believed to be currently in the hot phase,
\citep[e.g.,][]{Cio++91}. The exact phase space distribution of the
gas lost during stellar evolution depends on the complex interplay of
winds, inflows and outflows, cooling, AGN heating, interactions with
the environment and accretion of additional ``unprocessed'' gas
\citep[e.g.,][]{P+C98,Pip++05}. Determining the fate of the gas is
beyond the scope of this paper. However, X-ray observations show that
the residual gas is a small fraction of the stellar mass
\citep[e.g.][]{M+B03c,Hum++06} and therefore most of the gas must be
either expelled or recycled into secondary episodes of star
formation. For simplicity, in this analysis, we will consider two
extrema that should bracket the exact solution. In our default
scenario, all gas that is not recycled is dispersed and is therefore
counted by our two component model in the dark matter halo. In this
case, \mseld\, needs to be compared with the current \msesed
(including of course stellar remnants such as black holes and neutron
stars). In the other extreme, all gas lost retains the distribution
function of the stars and is therefore counted by the lensing and
dynamical two-component model in the stellar component. In this latter
case, which is effectively an upper limit to the dynamical importance
of residual gas, \mseld\, needs to be compared with \msesed\, at
zero-age.

\section{Results}
\label{sec:res}

\begin{figure}
\begin{center}
\resizebox{\columnwidth}{!}{\includegraphics{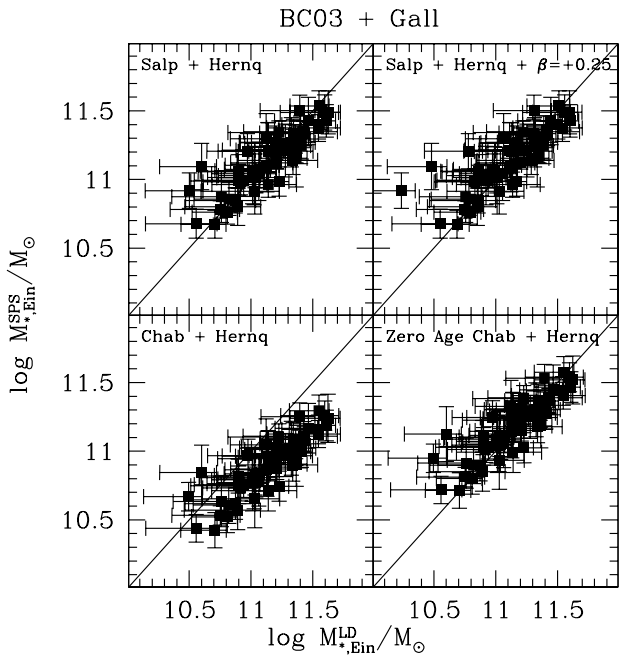}}
\end{center}
\figcaption{Comparison between stellar mass in the cylinder of
radius equal to the Einstein Radius as inferred from lensing and
dynamical models (x-axis) and that inferred from fitting stellar
populations synthesis models to the observed spectral energy
distribution (y-axis).  The solid line indicates the identity. Stellar
populations synthesis models by \citet{B+C03} are assumed together
with an informative metallicity prior \citep{Gal++05}.
\label{fig:comp}}
\end{figure}

\begin{figure}
\begin{center}
\resizebox{\columnwidth}{!}{\includegraphics{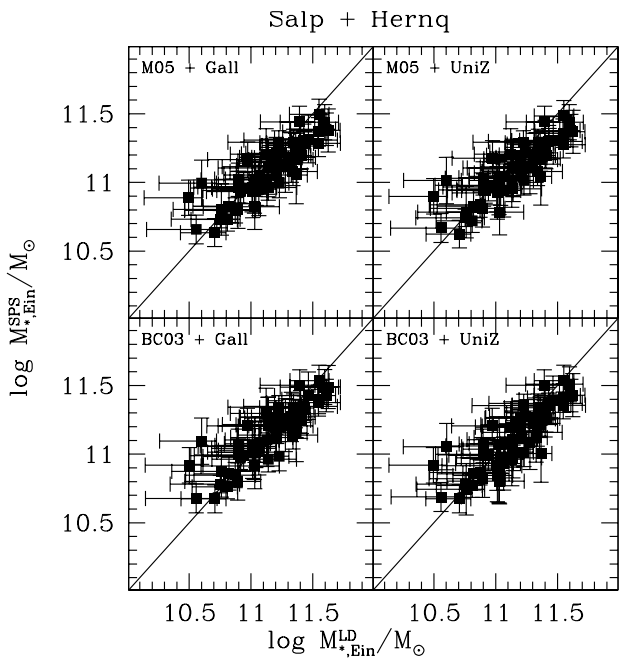}}
\end{center}
\figcaption{As in Figure~\ref{fig:comp} for different choices of stellar 
population synthesis models and metallicity priors. Salpeter IMF and
Hernquist stellar component models are assumed.
\label{fig:M05vsBC03}}
\end{figure}

A comparison of the two independent determinations of stellar mass
(\mseld\, and \msesed) is shown in Fig~\ref{fig:comp}, for four
combinations of IMF and lensing and dynamical models.  The two
quantities are tightly correlated, with scatter consistent with
observational errors. Notice that for a Salpeter IMF the points lie on
average around the identity line, and that changing anisotropy of
stellar orbits has very little effects on the inferred \mseld. For a
Chabrier IMF, in contrast, the current stellar mass underestimates
that inferred from stellar populations synthesis models. Finally, the
zero-age \msesed\, is larger than \mseld\, even for a Chabrier IMF.
The trends are robust with respect to the choice of stellar population
synthesis models or metallicity priors, as illustrated in
Figure~\ref{fig:M05vsBC03}.  We note that the data appear to suggest
that the relation between \mseld\, and \msesed\, is not linear: at low
masses the data appear to lie above the line indicating the identity,
while at high masses the data appear to lie below the identity. We
will return to this point in Section~\ref{ssec:alphaslope} after we
discuss in more detail the overall normalization in
Section~\ref{ssec:alpha}.

\subsection{Towards an absolute normalization of the IMF of massive galaxies: the ``IMF mismatch'' parameter}
\label{ssec:alpha}

\begin{figure}
\begin{center}
\resizebox{\columnwidth}{!}{\includegraphics{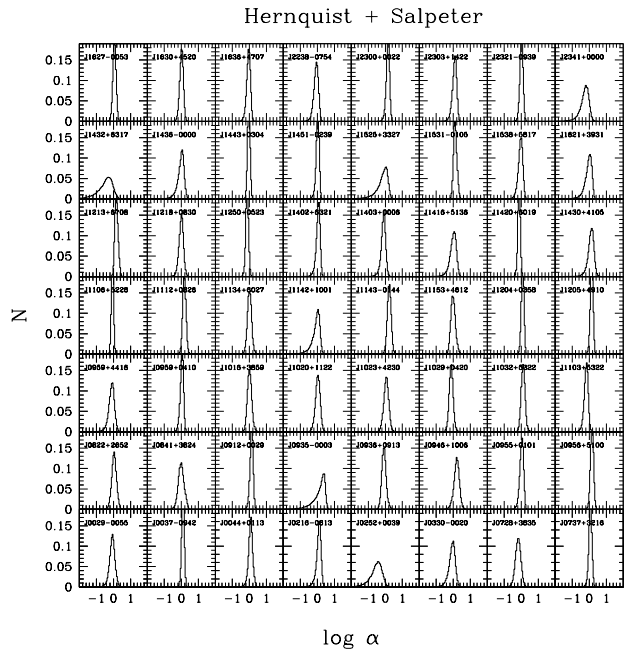}}
\end{center}
\figcaption{Posterior distribution function for the IMF mismatch parameter $\alpha\equiv$\mseld/\msesed with respect to a Salpeter IMF. 
\label{fig:histos}}
\end{figure}

The goal of this paper is to go beyond a comparison and determine the
absolute normalization of the IMF for each galaxy. This is used to
investigate its universality in terms of intrinsic scatter and
dependency on galaxy parameters such as velocity dispersion (hence
mass) and luminosity.  For this purpose, we introduce an ``IMF
mismatch'' parameter $\alpha \equiv${\mseld}/{\msesed}. For each
galaxy, we determine the posterior distribution function for $\alpha$
by combining samples drawn from the posterior distribution function
for \mseld\, and from that for \msesed. The resulting posterior
distribution samples for $\alpha$ assuming Salpeter IMF are shown in
Figure~\ref{fig:histos} for illustration. The median values of
$\alpha$ for each lens are given in Table~\ref{tab:alpha}, together
with other key properties of the lens galaxies, taken from
\citet{Aug++09a} and references therein.

The resulting average values of $\log \alpha$ for a variety of stellar
population synthesis models are summarized in~Table~\ref{tab:sum}. The
statistical uncertainty on $\langle \log \alpha \rangle$ is 0.03 dex
for any given model. The different choices of dynamical model
(Hernquist vs Jaffe, isotropic vs anisotropic) influence the average
of $\log \alpha$ only at a level of 0.02, which can then be neglected
in the rest of the discussion and considered as an additional
systematic uncertainty. Confirming the trends shown in
Figure~\ref{fig:comp}, the statistical analysis shows that a Salpeter
IMF tends to provide on average a much closer match between \msesed\,
and \mseld\, than a Chabrier IMF which appears to produce \msesed\,
that are systematically lower than \mseld\, (by $0.25\pm0.03\pm0.02$
dex).

Previous authors used comparisons between independent determinations
of stellar masses to select one IMF versus another. For example,
\citet{Gri++09} concluded that the Chabrier IMF underestimates stellar
masses, and therefore preferred a Salpeter IMF, based on \citet{B+C03}
models. However, \citet{Gri++09} only used the average dark matter
fraction as published for a subset of the SLACS lenses and therefore
could not make a detailed comparison for each individual case. We also
note that the stellar masses inferred by \citet{Gri++09} for the
galaxies in common with this study are systematically different than
the ones adopted here. As discussed by \cite{Aug++09a}, this
difference is due to a combination of photometry differences, and
choice of priors for metallicity and age parameters (e.g., Grillo et
al. 2009 assume constant solar metallicity). However the differences
are small and do not change the overall normalization, which is in
very good agreement between the two studies.

In contrast, \citet{Cap++06} used different stellar populations
synthesis models \citep{Vaz++96} than the ones adopted here and found
that stellar masses based on a Salpeter IMF were in some cases too
high compared to those determined with stellar kinematics, reaching
the opposite conclusion.  However -- in presence of significant
statistical errors -- this is actually expected, even if the Salpeter
IMF were a perfect match to the intrinsic IMF, given that measurements
will tend to scatter below and above the identity line. Indeed, even
with our own method, there are systems for which \msesed\, exceeds
\mseld\, for a Salpeter IMF, even though the average of the median
$\alpha$ is close to unity. Another caveat that must be kept in mind
when comparing to the \citet{Cap++06} study is that their sample
extends to significantly less massive galaxies than ours (velocity
dispersions $\sigma$ as low as 60 \kms, as opposed to our lower limit
of approximately 200 \kms).  Thus the two samples can only be compared
directly if the IMF does not depend on velocity dispersion or on
galaxy mass.

Regardless of the interpretation in terms of a specific IMF -- which
depends also on the uncertainties of the stellar populations models
\citep[e.g.][]{Mar05} -- we emphasize that our method provides an
absolute calibration of the stellar mass. The current stellar masses
given in Auger et al.\ (2009) assuming a Salpeter IMF multiplied by
$\alpha$ are {\it absolutely} calibrated against those inferred by the
lensing and dynamical models. On average, the \citet{Aug++09a}
Salpeter masses are calibrated to within 0.04 dex, even without
applying the IMF mismatch parameter. In general, the $\alpha$ values
given in~Table~\ref{tab:alpha} can be used to calibrate any stellar
population synthesis model, for any arbitrary choice of IMF, and
destiny of the gas lost during stellar evolution.  Remarkably, the
data are consistent with {\it very little intrinsic scatter} in $\log
\alpha$.  The upper limit on the intrinsic scatter is 0.09 dex (95\%
CL), i.e. {\it the absolute normalization of the IMF is uniform to
better than 25\%}.

\subsection{Universal or not? Trends with galaxy properties}
\label{ssec:alphaslope}

Within the class of massive early-type galaxies, the SLACS lenses span
approximately a factor of two in velocity dispersion and a factor of
ten in luminosity and stellar mass \citep{Aug++09a}. In turn, these
quantities correlate with the average epoch of formation of their
stellar populations, as well as their average metalliticity, abundance
ratios, and gas content
\citep[e.g.,][]{Tre++05b,Tho++05,Gal++05,Jim++07,Pip++09,GFS09}. 
If the IMF were to evolve during the epoch of formation of most of the
stars of early-type galaxies, or if it were to depend on the mode of
star formation or on the physical condition of the progenitor gas, we
would expect $\alpha$ to vary across our sample.

To test for signs of mass dependency of the IMF normalization, we
checked for a correlation between $\alpha$ and three indicators of
galaxy ``mass'': i) $\sigma_{\rm SIE}$, the velocity dispersion of the
best fitting lensing model \citep{Bol++08a,Aug++09a}; ii)
$\sigma_{*}$, the stellar velocity dispersion within the SDSS fiber
aperture; iii) the total V band luminosity corrected to a common
redshift $z=0.2$ as described by \citet{Aug++09a}.  The first two
choices are motivated by several lines of evidence
\citep[e.g.][]{GFS09} that indicate that velocity dispersion is the
most important parameter in determining stellar populations.  The
first quantity correlates well with stellar velocity dispersion and is
measured much more accurately \citep{Tre++06}. The errors on
$\sigma_{\rm SIE}$ are effectively negligible with respect to those on
stellar velocity dispersion, which dominate the error on \mseld. Thus,
this choice makes the covariance between $\alpha$ and $\sigma_{\rm
SIE}$ negligible.  The canonical stellar velocity dispersion
$\sigma_*$ suffers from a larger covariance with $\alpha$ due to its
larger errors. The third quantity, the V-band luminosity, is an
inferior galaxy mass proxy -- because it is sensitive to relatively
minor recent episodes of star formation -- and is inversely covariant
with $\alpha$ because to first order \msesed is proportional to
L$_{\rm V}$.

\begin{figure*}
\begin{center}
\resizebox{0.3\textwidth}{!}{\includegraphics{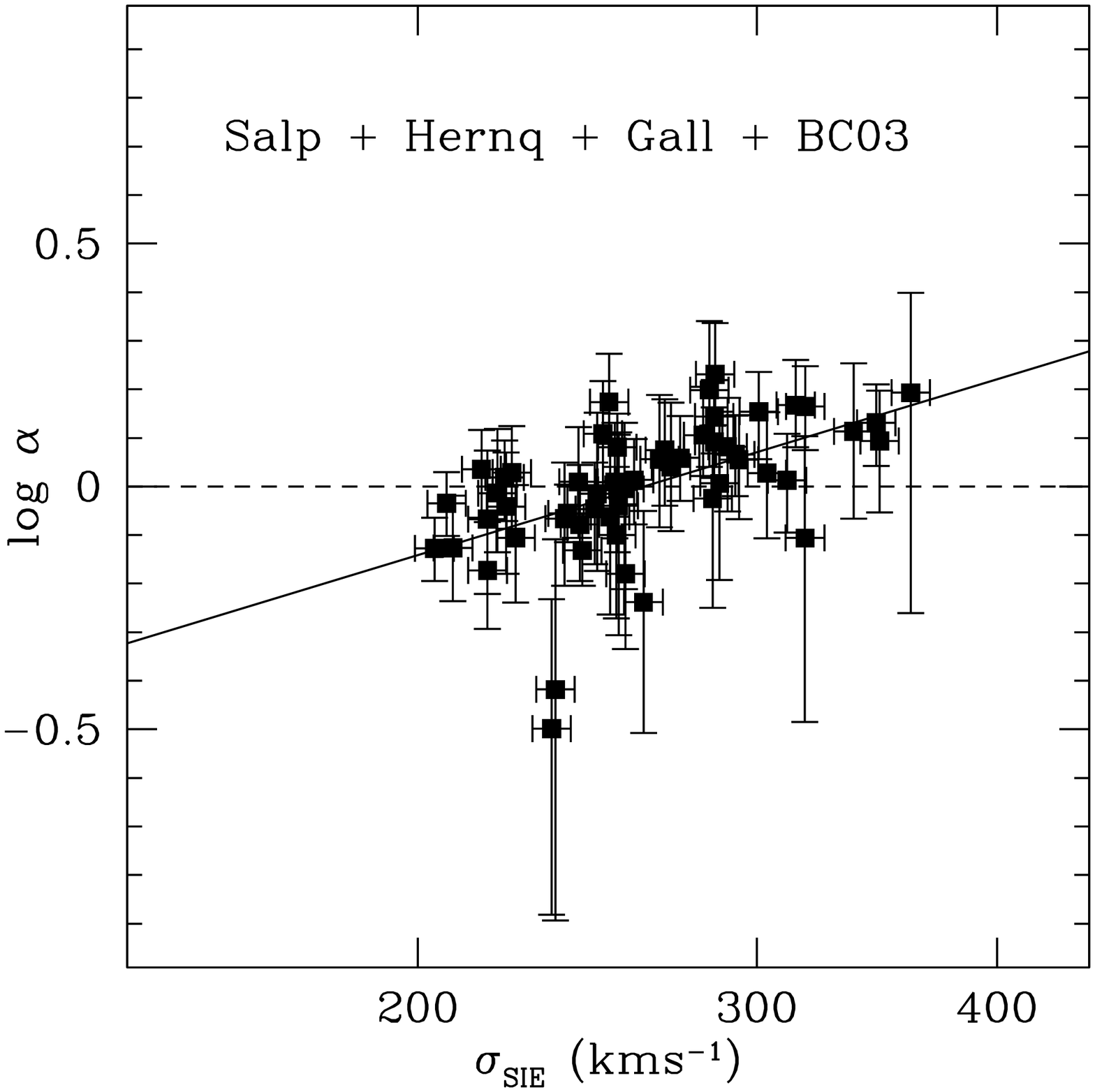}}
\resizebox{0.3\textwidth}{!}{\includegraphics{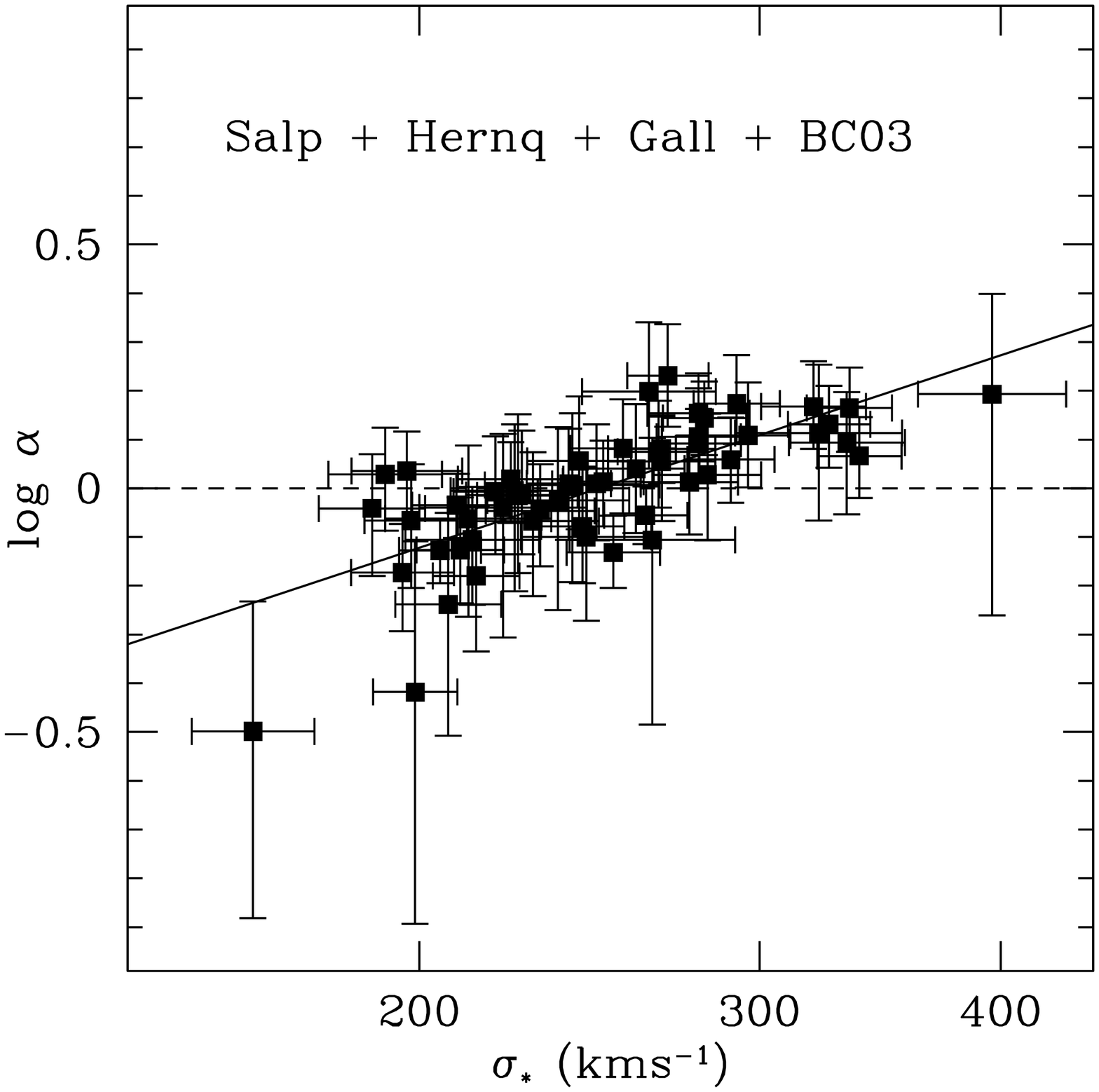}}
\resizebox{0.3\textwidth}{!}{\includegraphics{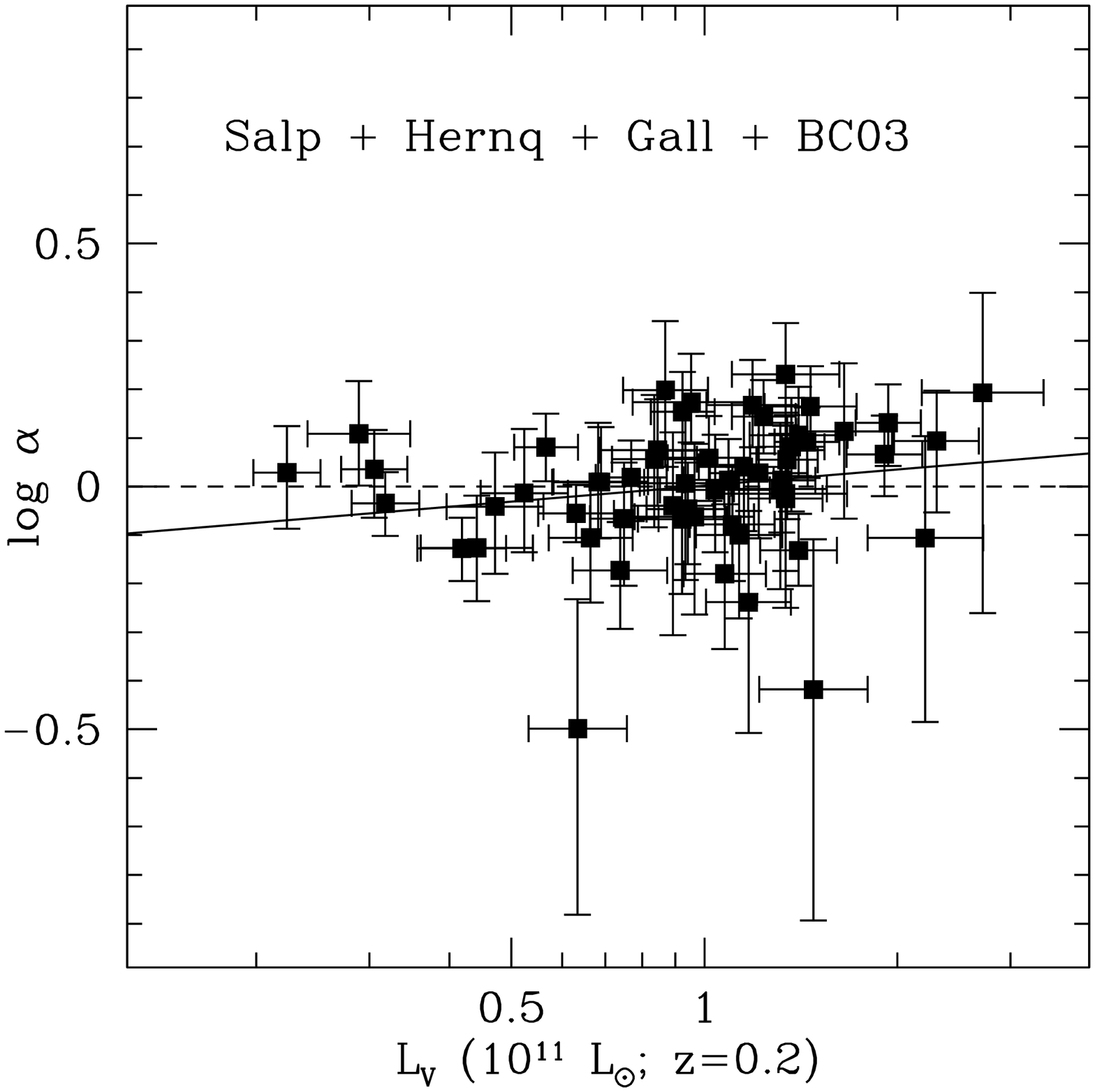}}
\end{center}
\figcaption{Template mismatch parameter
$\alpha\equiv${\mseld}/{\msesed} for Salpeter IMF as a function of
lensing velocity dispersion (left), stellar velocity dispersion
(center) and V-band luminosity corrected to $z=0.2$.  A tentative
positive trend with velocity dispersion is observed (solid line). The
dashed line represents the trend expected for a universal Salpeter IMF.
\label{fig:ptrends}}
\end{figure*}

The results are shown in Figure~\ref{fig:ptrends} using the Salpeter
IMF as baseline and for our standard stellar populations models
(adopting a Chabrier IMF would move all the points upwards by $\sim$
0.25 dex, while all the other choices introduce negligible changes). A
trend with non-zero slope is detected for $\sigma_{\rm SIE}$ and
$\sigma_*$. No significant slope is found for L$_{\rm V}$.  The best
fit linear relations are found to be, with no evidence of intrinsic
scatter:

\begin{equation}
\log \alpha = (1.20\pm0.25) \log \sigma_{\rm SIE} - 2.91\pm0.02,
\label{eq:trends1}
\end{equation}

\begin{equation}
\log \alpha = (1.31\pm0.16) \log \sigma_{*} - 3.14\pm0.01,
\label{eq:trends2}
\end{equation}

\begin{equation}
\log \alpha = (0.11\pm0.08) \log {\rm L}_{\rm V} + 0.00 \pm0.02,
\label{eq:trends3}
\end{equation}

\noindent
where velocity dispersions are in units of \kms\, and L$_{\rm V}$ is in
units of 10$^{11}$ L$_{\odot,\rm V}$.

In summary, $\log \alpha $ appears to increase with velocity
dispersion by an amount comparable to the difference between a
Chabrier and Salpeter IMF over the range probed. The slope of the
correlation with luminosity is significantly smaller then expected
given the correlations with velocity dispersion and the Faber Jackson
relation. The inverse covariance mentioned above is not sufficient to
explain the discrepancy unless there is significant intrinsic
scatter. This may therefore be another indication that velocity
dispersion and not luminosity is the main parameter controlling
stellar populations, including the IMF \citep{Ber++07,GFS09}.

\section{Discussion}
\label{sec:disc}

Before attempting to interpret our perhaps surprising findings, it is
important to emphasize a number of caveats: i) the sample is
relatively small and with a selection function that strongly favors
high velocity dispersion galaxies; ii) the quantities on the two axis
of Figure~\ref{fig:ptrends} are not independent, even though the known
errors are small enough for covariance not to be causing the observed
trends; iii) the stellar masses are typically based on three-band
photometry and therefore we can only probe simple star formation
histories. For all these reasons the results reported here must be
considered as tentative until verified by larger samples, spanning a
broader range of properties, and with the help of spectroscopic
diagnostics of stellar populations.

Keeping these caveats in mind, we now discuss the two main results of
this paper.  The first result is that the average absolute
normalization of the IMF inferred by our study is higher than those
commonly assumed when deriving masses for distant galaxies using
colors. Those ``lighter'' IMFs have been preferred on the basis of
studies of local stellar populations \citep[e.g.][]{Kro01,Cha03} as
well as on the basis of dynamical arguments applied to spiral galaxies
\citep[e.g.][]{B+d01}. However, those measurements do not necessarily 
apply to the stellar populations of massive early-type galaxies, if
the IMF is not universal, but depends, for example, on metallicity or
other conditions that vary with cosmic time \citep[e.g.][and
references therein]{Elm08}. The second result is the trend in IMF
normalization with galaxy velocity dispersion. Taken at face value,
this trend would imply that whereas a ``light'' IMF such as Chabrier's
is appropriate for systems with $\sigma\sim200$ \kms\, (and therefore
more or less consistent with the standard conclusions for
Milky-Way-type galaxies and spirals), for the most massive systems
there is a higher abundance of low mass stars.  This appears to be
quite a dramatic change over a factor of $\sim 2$ in velocity
dispersion, corresponding approximately to 0.15 dex change in [Fe/H]
and 0.1 dex in [$\alpha$/Fe] \citep{Ber++06c}. Part of this trend
could also be ascribed to larger fraction of gas loss during stellar
evolution being retained by higher velocity dispersion
systems. However, the amount of retained gas would have to be
comparable to the stellar mass in the central regions in order to
explain the trend completely, which seems to be ruled out by X-ray
observations \citep[e.g.][]{M+B03c}. Deep X-ray observations of the
SLACS sample would be useful to verify exactly how much gas is left.
More and better spectroscopic data are needed to investigate whether
this trend is real, or whether there are unknown systematic effects at
play. If the trend were to be confirmed, it would have far reaching
implications for the determination of the evolving mass function of
galaxies, because it may change its shape as well as its overall
normalization.

Alternatively -- if the IMF normalization were indeed universal over
the mass range spanned by the SLACS sample -- our finding would imply
that one of our assumptions in deriving \mseld\, is not warranted. As
we have shown, anisotropy or changes in the assumed stellar mass
density profile are not going to be sufficient, as they only change
$\alpha$ by a few hundredths of a dex at most. Only changing our
assumed dark matter density profile systematically with $\sigma$ would
have a sufficiently large effect to affect the observed trend. The
assumed break radius (and hence concentration parameter) has only
minimal effects. The dominant parameter in determining $f_*$ is the
inner slope of the dark matter density profile $\gamma$: steeper halos
require a smaller stellar mass fraction \citep[see,
e.g.,][]{T+K02a,K+T03,T+K04}. Thus -- in order to keep $\alpha$
constant and consistent with a Chabrier normalization -- dark matter
halos would need to be NFW-like at the low end of the velocity
dispersion range and become steeper towards the high end.  In contrast
-- in order to keep $\alpha$ constant and consistent with a Salpeter
normalization -- dark matter halos would need to be NFW-like at the
high end of the velocity dispersion range and become flatter towards
the low end, tending towards an inner constant core.

What could cause the dark matter inner slope to steepen with velocity
dispersion? Baryons are the primary suspect, since this trend is not
observed in dark matter only simulations. Baryons are indeed dominant
over this radial range, and they could be responsible for making the
dark matter density profile steeper than NFW
\citep[e.g.][]{Gne++04,J+K07}.  This baryonic compression would need
to be more effective for higher velocity dispersion objects in order
to explain the observed trend
\footnote{While this manuscript was under revision, we became aware of
the work by \citet{SMP09} who find a similar trend based on a joint
weak lensing and dynamical of early-type galaxies selected from the
SDSS survey.}.
By steepening the dark matter density profile, baryons would also
effectively increase the dark matter fraction within a fixed
aperture. This in turn would imply a correlation between dark matter
fraction and velocity dispersion.

\section{Conclusions}
\label{sec:conc}

We combined three independent diagnostics of mass (lensing, dynamics
and stellar populations synthesis models) to determine an absolute
normalization of the IMF for a sample of 56 early-type galaxies,
spanning over a decade in stellar mass and a factor of two in velocity
dispersion, under the assumption of a NFW dark matter density profile.
On average, the absolute IMF normalization is found to be close to
that of a Salpeter IMF and larger than that for a Chabrier IMF. Using
the prescription outlined in this paper, stellar masses based on any
stellar population synthesis models can be absolutely calibrated to
better than 20\%, a significant progress with respect to the range of
a factor of $\sim$2 spanned by standard choices of the IMF.

A tentative trend of IMF mismatch parameter $\alpha$=\mseld/\msesed
with galaxy velocity dispersion is found. Two possible explanations,
not necessarily mutually exclusive, are suggested for the observed
trend:

\begin{itemize}

\item The IMF is not universal, but rather depends on parameters 
such as metallicity, age, and abundance ratios of the stellar
populations. In order to fully explain the observed trend, the
normalization of the IMF and thus the abundance of low mass stars,
must increase from Chabrier-like for $\sigma~\sim 200$~\kms\, to
Salpeter-like for the most massive early-type galaxies.

\item Dark matter halos are not universal. For a uniform Chabrier IMF,
the observed trend of $\alpha$ with $\sigma$ could be explained if the
inner slope of the dark matter halo were systematically steeper than
NFW for the high velocity dispersion systems. For a uniform Salpeter
IMF, the the dark matter halos would have to be NFW at the high mass
end and flatter at lower masses.

\end{itemize}

In conclusion, the data are inconsistent with both a universal IMF and
universal NFW dark matter halos over the mass range probed by the
SLACS sample. There is a fundamental degeneracy between the two
interpretations that cannot be broken with the current dataset.

Further tests and more work are required to verify and extend our
perhaps surprising results. Firstly, we need to extend our samples to
cover a wider range of redshifts, masses, and morphological types and
thus probe a larger variety of IMFs. Secondly, we need to use spectral
stellar population diagnostics to obtain independent constraints on
the stellar mass to light ratios as well as on the physical parameters
that may correlate with IMF normalization. Thirdly, we need to improve
the constraints on the inner slope of the dark matter halo as a
function of velocity dispersion to break the current degeneracy. At
the level of individual galaxies, some progress can be achieved by
constructing more sophisticated lensing and dynamical models
\citep[e.g.,][]{Bar++09}. In fact, the models presented here only use
a single measurement of stellar velocity dispersion from SDSS and the
total mass enclosed by the Einstein Radius, while more radial
information can be extracted from both diagnostics. At the level of
joint analysis of sub samples of galaxies, we need to have enough
objects so that they can be binned by velocity dispersions to perform
a weak lensing analysis. The addition of weak lensing to the strong
lensing, dynamics, and stellar populations diagnostics would allow us
to probe systematic variations with velocity dispersion of the dark
matter halo shape and of the stellar to virial mass to light ratio
. If the amount of contraction is an important ingredient of the
observed trend with velocity dispersion, we expect to see a parallel
trend in the overall efficiency of converting baryons into stars, or
perhaps in the spatial concentration of the stellar component relative
to that of the halo.

\medskip

\acknowledgments

We thank L.~Bildsten, K.~Bundy, L.~Ciotti, M.~Cappellari, C.~Maraston,
L.~Moustakas, C.~Nipoti, S.~Pellegrini for many insightful comments
and stimulating conversations.  Support for programs \#10174, \#10587,
\#10886, \#10494, \#10798, \#11202 was provided by NASA through a
grant from the Space Telescope Science Institute, which is operated by
the Association of Universities for Research in Astronomy, Inc., under
NASA contract NAS 5-26555.  T.T.  acknowledges support from the NSF
through CAREER award NSF-0642621, by the Sloan Foundation through a
Sloan Research Fellowship, and by the Packard Foundation through a
Packard Fellowship.  L.V.E.K. is supported by an NWO-VIDI program
subsidy (project number 639.042.505).  PJM was given support by the
TABASGO foundation in the form of a research fellowship.

\bibliographystyle{apj}


\begin{deluxetable}{lccccc}
\setlength{\tabcolsep}{3pt}
\tablecaption{Basic parameters of the lens galaxies}
\tablehead{
\colhead{ID}        & 
\colhead{$z_l$}        &
\colhead{$\sigma_*$}        &
\colhead{$\sigma_{\rm SIE}$} & 
\colhead{L$_{\rm V}$ (z=0.2)} &
\colhead{$\log \alpha$} \\
\colhead{} &
\colhead{} &
\colhead{(\kms)} &
\colhead{(\kms)} &
\colhead{(10$^{11}$ M$_\odot$)} &
\colhead{}}
\tablecolumns{6}
\startdata
J0029-0055 & 0.2270 & 229$\pm$18 & 217.3$\pm$5.0 & 0.69$\pm$0.05 & -0.07$^{+0.14}_{-0.15}$ \\ 
J0037-0942 & 0.1955 & 279$\pm$14 & 285.2$\pm$6.6 & 1.09$\pm$0.06 &  0.09$^{+0.07}_{-0.07}$ \\ 
J0044+0113 & 0.1196 & 266$\pm$13 & 268.7$\pm$6.2 & 0.62$\pm$0.06 &  0.08$^{+0.11}_{-0.11}$ \\ 
J0216-0813 & 0.3317 & 333$\pm$23 & 347.6$\pm$8.0 & 1.81$\pm$0.12 &  0.09$^{+0.10}_{-0.15}$ \\ 
J0252+0039 & 0.2803 & 164$\pm$12 & 234.7$\pm$5.4 & 0.49$\pm$0.04 & -0.50$^{+0.26}_{-0.38}$ \\ 
J0330-0020 & 0.3507 & 212$\pm$21 & 251.7$\pm$5.8 & 0.74$\pm$0.07 & -0.06$^{+0.15}_{-0.20}$ \\ 
J0728+3835 & 0.2058 & 214$\pm$11 & 256.4$\pm$5.9 & 0.82$\pm$0.05 & -0.18$^{+0.14}_{-0.16}$ \\ 
J0737+3216 & 0.3223 & 338$\pm$17 & 292.5$\pm$6.7 & 1.51$\pm$0.09 &  0.07$^{+0.08}_{-0.09}$ \\ 
J0822+2652 & 0.2414 & 259$\pm$15 & 270.8$\pm$6.2 & 0.87$\pm$0.06 &  0.04$^{+0.13}_{-0.13}$ \\ 
J0841+3824 & 0.1159 & 225$\pm$11 & 247.9$\pm$5.7 & 0.96$\pm$0.09 & -0.01$^{+0.17}_{-0.16}$ \\ 
J0912+0029 & 0.1642 & 326$\pm$16 & 346.2$\pm$8.0 & 1.47$\pm$0.07 &  0.13$^{+0.08}_{-0.09}$ \\ 
J0935-0003 & 0.3475 & 396$\pm$35 & 360.8$\pm$8.3 & 2.06$\pm$0.20 &  0.19$^{+0.20}_{-0.44}$ \\ 
J0936+0913 & 0.1897 & 243$\pm$12 & 242.8$\pm$5.6 & 0.83$\pm$0.06 & -0.08$^{+0.12}_{-0.12}$ \\ 
J0946+1006 & 0.2219 & 263$\pm$21 & 283.5$\pm$6.5 & 0.66$\pm$0.04 &  0.20$^{+0.14}_{-0.16}$ \\ 
J0955+0101 & 0.1109 & 192$\pm$13 & 223.8$\pm$5.1 & 0.17$\pm$0.01 &  0.03$^{+0.10}_{-0.12}$ \\ 
J0956+5100 & 0.2405 & 334$\pm$17 & 318.0$\pm$7.3 & 1.12$\pm$0.08 &  0.16$^{+0.08}_{-0.09}$ \\ 
J0959+4416 & 0.2369 & 244$\pm$19 & 253.6$\pm$5.8 & 0.86$\pm$0.06 & -0.10$^{+0.14}_{-0.17}$ \\ 
J0959+0410 & 0.1260 & 197$\pm$13 & 215.8$\pm$5.0 & 0.23$\pm$0.01 &  0.04$^{+0.08}_{-0.10}$ \\ 
J1016+3859 & 0.1679 & 247$\pm$13 & 253.2$\pm$5.8 & 0.51$\pm$0.04 &  0.01$^{+0.12}_{-0.12}$ \\ 
J1020+1122 & 0.2822 & 282$\pm$18 & 303.7$\pm$7.0 & 0.94$\pm$0.07 &  0.03$^{+0.13}_{-0.13}$ \\ 
J1023+4230 & 0.1912 & 242$\pm$15 & 267.1$\pm$6.1 & 0.63$\pm$0.04 &  0.06$^{+0.13}_{-0.14}$ \\ 
J1029+0420 & 0.1045 & 210$\pm$11 & 208.6$\pm$4.8 & 0.33$\pm$0.03 & -0.13$^{+0.11}_{-0.11}$ \\ 
J1032+5322 & 0.1334 & 296$\pm$15 & 249.6$\pm$5.7 & 0.22$\pm$0.02 &  0.11$^{+0.11}_{-0.11}$ \\ 
J1103+5322 & 0.1582 & 196$\pm$12 & 217.4$\pm$5.0 & 0.56$\pm$0.04 & -0.17$^{+0.10}_{-0.12}$ \\ 
J1106+5228 & 0.0955 & 262$\pm$13 & 239.2$\pm$5.5 & 0.47$\pm$0.03 & -0.06$^{+0.06}_{-0.06}$ \\ 
J1112+0826 & 0.2730 & 320$\pm$20 & 314.4$\pm$7.2 & 0.90$\pm$0.06 &  0.17$^{+0.09}_{-0.09}$ \\ 
J1134+6027 & 0.1528 & 239$\pm$12 & 242.4$\pm$5.6 & 0.52$\pm$0.04 &  0.01$^{+0.11}_{-0.12}$ \\ 
J1142+1001 & 0.2218 & 221$\pm$22 & 254.3$\pm$5.8 & 0.68$\pm$0.04 & -0.04$^{+0.15}_{-0.27}$ \\ 
J1143-0144 & 0.1060 & 269$\pm$13 & 285.5$\pm$6.6 & 0.94$\pm$0.08 &  0.23$^{+0.11}_{-0.10}$ \\ 
J1153+4612 & 0.1797 & 226$\pm$15 & 220.0$\pm$5.1 & 0.39$\pm$0.03 & -0.01$^{+0.13}_{-0.12}$ \\ 
J1204+0358 & 0.1644 & 267$\pm$17 & 253.9$\pm$5.8 & 0.44$\pm$0.02 &  0.08$^{+0.07}_{-0.07}$ \\ 
J1205+4910 & 0.2150 & 281$\pm$14 & 285.2$\pm$6.6 & 0.96$\pm$0.05 &  0.14$^{+0.07}_{-0.08}$ \\ 
J1213+6708 & 0.1229 & 292$\pm$15 & 251.4$\pm$5.8 & 0.68$\pm$0.06 &  0.17$^{+0.10}_{-0.08}$ \\ 
J1218+0830 & 0.1350 & 219$\pm$11 & 254.0$\pm$5.8 & 0.78$\pm$0.07 & -0.01$^{+0.11}_{-0.13}$ \\ 
J1250+0523 & 0.2318 & 252$\pm$14 & 243.5$\pm$5.6 & 1.08$\pm$0.06 & -0.13$^{+0.08}_{-0.07}$ \\ 
J1402+6321 & 0.2046 & 267$\pm$17 & 293.7$\pm$6.8 & 1.05$\pm$0.05 &  0.05$^{+0.09}_{-0.12}$ \\ 
J1403+0006 & 0.1888 & 213$\pm$17 & 224.8$\pm$5.2 & 0.50$\pm$0.03 & -0.11$^{+0.11}_{-0.13}$ \\ 
J1416+5136 & 0.2987 & 240$\pm$25 & 287.0$\pm$6.6 & 0.73$\pm$0.05 &  0.01$^{+0.15}_{-0.20}$ \\ 
J1420+6019 & 0.0629 & 205$\pm$10 & 204.0$\pm$4.7 & 0.31$\pm$0.02 & -0.13$^{+0.06}_{-0.07}$ \\ 
J1430+4105 & 0.2850 & 322$\pm$32 & 336.9$\pm$7.7 & 1.28$\pm$0.10 &  0.11$^{+0.14}_{-0.18}$ \\ 
J1432+6317 & 0.1230 & 199$\pm$10 & 235.8$\pm$5.4 & 1.10$\pm$0.09 & -0.41$^{+0.31}_{-0.48}$ \\ 
J1436-0000 & 0.2852 & 224$\pm$17 & 256.1$\pm$5.9 & 1.00$\pm$0.07 & -0.01$^{+0.14}_{-0.21}$ \\ 
J1443+0304 & 0.1338 & 209$\pm$11 & 207.0$\pm$4.8 & 0.24$\pm$0.01 & -0.03$^{+0.06}_{-0.07}$ \\ 
J1451-0239 & 0.1254 & 223$\pm$14 & 221.9$\pm$5.1 & 0.57$\pm$0.03 &  0.02$^{+0.08}_{-0.09}$ \\ 
J1525+3327 & 0.3583 & 264$\pm$26 & 317.9$\pm$7.3 & 1.72$\pm$0.16 & -0.11$^{+0.21}_{-0.37}$ \\ 
J1531-0105 & 0.1596 & 279$\pm$14 & 281.4$\pm$6.5 & 1.02$\pm$0.08 &  0.11$^{+0.10}_{-0.08}$ \\ 
J1538+5817 & 0.1428 & 189$\pm$12 & 222.3$\pm$5.1 & 0.36$\pm$0.03 & -0.04$^{+0.11}_{-0.14}$ \\ 
J1621+3931 & 0.2449 & 236$\pm$20 & 284.6$\pm$6.5 & 1.03$\pm$0.06 & -0.02$^{+0.15}_{-0.22}$ \\ 
J1627-0053 & 0.2076 & 290$\pm$15 & 273.8$\pm$6.3 & 0.79$\pm$0.05 &  0.06$^{+0.09}_{-0.09}$ \\ 
J1630+4520 & 0.2479 & 276$\pm$16 & 311.1$\pm$7.2 & 1.04$\pm$0.05 &  0.01$^{+0.10}_{-0.11}$ \\ 
J1636+4707 & 0.2282 & 231$\pm$15 & 247.2$\pm$5.7 & 0.73$\pm$0.05 & -0.05$^{+0.10}_{-0.11}$ \\ 
J2238-0754 & 0.1371 & 198$\pm$11 & 238.4$\pm$5.5 & 0.56$\pm$0.03 & -0.07$^{+0.12}_{-0.14}$ \\ 
J2300+0022 & 0.2285 & 279$\pm$17 & 300.8$\pm$6.9 & 0.72$\pm$0.04 &  0.15$^{+0.08}_{-0.10}$ \\ 
J2303+1422 & 0.1553 & 255$\pm$16 & 289.7$\pm$6.7 & 1.03$\pm$0.05 &  0.08$^{+0.10}_{-0.13}$ \\ 
J2321-0939 & 0.0819 & 249$\pm$12 & 259.1$\pm$6.0 & 0.78$\pm$0.05 &  0.01$^{+0.08}_{-0.09}$ \\ 
J2341+0000 & 0.1860 & 207$\pm$13 & 262.0$\pm$6.0 & 0.89$\pm$0.06 & -0.24$^{+0.19}_{-0.26}$ \\
\enddata
\label{tab:alpha}
\tablecomments{The IMF mismatch parameter $\alpha$ is given with
respect to a Salpeter IMF assuming BC03 models and Gallazzi
metallicity prior. $\sigma_*$ is the SDSS-measured stellar velocity
dispersion within the spectroscopic aperture, as given in SLACS paper
V \citep{Bol++08a}.}
\end{deluxetable}

\begin{deluxetable}{lllc}
\setlength{\tabcolsep}{3pt}
\tabletypesize{\scriptsize}
\tablecaption{Average IMF mismatch parameter ($\langle \log \alpha \rangle$)}
\tablehead{
\colhead{IMF}        & 
\colhead{SSP models}        &
\colhead{Z Prior}        &
\colhead{$\langle \log \alpha \rangle$}}
\tablecolumns{4}
\startdata
Salpeter & BC03 & Gallazzi & 0.00 \\
Chabrier & BC03 & Gallazzi & 0.25 \\ 
Salpeter & BC03 & Uniform  & 0.03 \\ 
Chabrier & BC03 & Uniform  & 0.27 \\
Salpeter & M05  & Gallazzi & 0.05 \\
Salpeter & M05  & Uniform  & 0.06 \\
\enddata
\label{tab:sum}
\tablecomments{Statistical errors are 0.03 dex.}
\end{deluxetable}

\end{document}